\documentstyle[aps,prb,psfig,floats]{revtex}
\begin{document}
\draft
\twocolumn[\hsize\textwidth\columnwidth\hsize\csname @twocolumnfalse\endcsname
\title{
{Stable and Metastable Structures of Cobalt on Cu(001): An {\em ab initio} Study} }
\author{R.~Pentcheva and M.~Scheffler}
\address{
{ Fritz-Haber-Institut der Max-Planck-Gesellschaft, Faradayweg 4-6, D-14195
  Berlin-Dahlem, Germany }
}
\date{Received 12 August 1999}
\maketitle
\begin{abstract}
We report results of density-functional theory calculations on the structural,
magnetic, and electronic properties of $(1\times 1)$-structures of Co on
Cu(001) for coverages up to two monolayers. In particular we discuss the
tendency towards phase separation in Co islands and the possibility of
segregation of Cu on top of the Co-film. A sandwich structure consisting of a
bilayer Co-film covered by 1ML of Cu is found to be the lowest-energy
configuration. We also discuss a bilayer c$(2\times 2)$-alloy which may form
due to kinetic reasons, or be stabilized at strained surface
regions. Furthermore, we study the influence of magnetism on the various
structures and, e.g., find that Co adlayers induce a weak spin-density wave in 
the copper substrate. 
           
\end{abstract}

]
\narrowtext
%
\section{Introduction}\label{intro}
Heteroepitaxial structures of Co and Cu exhibit intriguing
magnetic properties such as giant magnetoresistance \cite{BGSZ89},
interlayer exchange coupling \cite{Bru95}, and surface magnetic anisotropy
\cite{BC91}. Since these properties are closely related to 
the surface and interface morphology, identification and understanding of the 
atomic structures and energetics of the adsorption of cobalt on the copper
surface are of great interest. Specifically we discuss in this paper the
[001]-surface orientation. Thin films deposited on a substrate of a different
material are generally subject to strain arising from the different lattice
parameters of the adsorbate and substrate.  Our calculations show that the
lattice constant of a ferromagnetic fcc bulk phase of Co is
$2.8\%$ smaller than that of a fcc Cu crystal, while the lattice
constant of a hypothetical nonmagnetic fcc cobalt crystal is $4.3\%$ smaller
than that of the copper crystal. Here we take the fcc structure of cobalt,
because it has been shown that a thick epitaxial cobalt film on Cu(001)
can be characterized in terms of a tetragonally distorted face centered
cubic (fct) phase~\cite{papm97}. The lattice mismatch between cobalt and
copper suggests a small tensile strain. However for ultrathin films ($\Theta <
2$ ML) the comparison of the bulk phases of adsorbate and substrate is not
necessarily  very relevant. For example total-energy
calculations~\cite{dipl96} show that the equilibrium lattice constant
of an unsupported Co monolayer is $14.1\%$ (nonmagnetic case)
and $12.2\%$ (ferromagnetic case) smaller than the Cu bulk lattice constant,
implying that an ultrathin film might be subject to a much stronger tensile
strain than a thick overlayer. The relation between lattice mismatch and
relaxation of the interlayer spacing will be discussed in Section
\ref{struct} below.

While experimental studies of coverages above 2ML show that growth
proceeds in an almost perfect layer-by-layer mode, for the initial two layers
a deviation from  the Frank-van der Merwe (FM) growth mode and a strong
dependence on the growth conditions was reported~\cite{fassb}. 
Angle-resolved X-ray photoemission spectroscopy (ARXPS) data \cite{lit}
indicate that the second layer begins to form before the first layer is
completed. Assuming the coexistence of areas of clean Cu(001) surface and of
monolayer and bilayer islands, a LEED analysis~\cite{kir1} estimated that for
a total coverage of one monolayer and deposition rates ranging between 0.016
and 0.33 ML/s the area covered by bilayer islands at room temperature is
$20-40\%$. Fassbender \emph{et al.}\cite{fassb} performed STM-experiments at
room temperature for a total coverage of 1.35~ML and report that the fractional
layer filling depends strongly on the deposition rate. For a low (0.003 ML/s)
deposition rate they found that the first layer was closed and 0.35ML were in
the second layer, while for a high deposition rate (0.3 ML/s) $15\%$ of the
surface was still uncovered and about $50\%$ of the surface was already
covered by bilayer high islands.

X-ray photoemission scattering (XPS)\cite{lit}, Auger
electron scattering(AES) and STM measurements\cite{kir2} show an increase
of the Cu signal and decrease of the Co signal upon annealing which was
interpreted as segregation of substrate material on top of the cobalt layer. 
Similar results were reported for Fe/Cu(001)\cite{kir4}. This effect was
explained in terms of the lower surface energy of Cu compared to Co. We note
that the application of this argument to thin film systems is not trivial
because of the energy cost of the additionally created Cu/Co interface. Yet
our studies show that in the case of Co on Cu(001) the contribution of the 
interface energy is very small (see Section~\ref{stabil}). 

The impact of morphological changes on the magnetic properties of
Co/Cu(001) was recently investigated with X-ray magnetic circular dichroism
(XMCD) and magneto optical Kerr effect (MOKE)~\cite{babs0} experiments. At a
cobalt coverage of $1.8$~ML a sudden jump of the Curie temperature was 
measured which changed strongly with time or a subsequent heat treatment. The
authors speculated that the critical thickness coincides with the thickness at
which bilayer cobalt islands coalesce.

Depending on growth conditions (temperature, deposition rate), significantly
different structures are observed experimentally. Although the magnetic
properties of Co on Cu(001) have been the subject of many theoretical studies,
a systematic theoretical analysis of the different configurations and  their
relative stability is still lacking. Moreover most of the
calculations~\cite{sb2,lmto2} have used slabs with atomic positions frozen to
the bulk coordinates of the substrate, neglecting thus the structural
relaxation of the clean Cu(001) surface and of the Co/Cu(001) adsorbate system.

In this paper we focus on the behavior of Co on Cu(001)
under thermodynamic equilibrium conditions. We
performed density-functional theory calculations considering a variety of
configurations ($\theta\leq2$~ML). In particular we discuss two
aspects: the formation of multilayer cobalt islands and sandwich structures
with a copper capping. For each system we performed a full structure
optimization and establish the relation between the energetic trends and the
structural, magnetic and electronic properties.
 The paper is organized as follows: The details of the calculations
 are given in Section \ref{theory}. In Section~\ref{stabil} we discuss the
stability of the systems against separation in multilayer islands and the
influence of the capping layer. The structural
(Section~\ref{struct}), magnetic (Section~\ref{magn}) and electronic
(Section~\ref{electr}) properties of mono- and bilayer cobalt films on
Cu(001), as well as of the corresponding copper capped systems are
investigated. Finally in Section \ref{co111_co001} we address the
similarities  and differences between Co/Cu(001) and Co/Cu(111) referring to
STM and  {\em ab initio} results for the
$[111]$-orientation\cite{norskov}. The results are summarized in Section
\ref{Summary}. 
\section{Calculational Details }\label{theory}
Our calculations are performed using density-functional theory (DFT). The
exchange-correlation functional is treated within the local-density
approximation (LDA)~\cite{pw} and for the magnetic systems we
performed spin-polarized calculations within the  local spin-density
approximation (LSDA). We also examined the possible importance of non-local
exchange correlation effects by employing the generalized-gradient
approximation (GGA) in the parameterization of Perdew, Burke, and
Ernzerhof~\cite{pbe96}. The results show that for our study LDA and GGA give
the same structural and energetic trends. More details on this issue will be
discussed in the Appendix. 

The Kohn-Sham equation was solved applying the full-potential 
linearized augmented plane wave (FP-LAPW) method \cite{Bla93,max}. The surface
is simulated by repeated slabs separated in  $z$-direction
by a vacuum region. Co is adsorbed on both sides of the substrate. The
thickness of the vacuum region between the slabs, corresponding to 6 Cu layers
($10.65$~{\AA}), is found to be sufficient to avoid interactions of
the Co atoms. The interlayer distances $d_{\rm 12}$ and
$d_{\rm 23}$ were optimized with a damped Newton dynamics and the relaxations
$\Delta d_{\rm 12}/d_0$ and $\Delta d_{\rm 23}/d_0$ are given with respect
to the interlayer spacing of a Cu crystal, ${d_0}$. Referring the Co-Cu and
for Co bilayer systems even the Co-Co interlayer distances to the interlayer
spacing of Cu is probably not an optimum choice, but it is well defined and
has been the common practice  for such adsorbate systems. We therefore use
this convention here as well.

The lattice constant for the fcc copper crystal $a_{\rm Cu} = 3.55$~{\AA},
obtained from a non-relativistic calculation,  is $1.6\%$ smaller than the
measured one ($3.61$~{\AA}), $0.1\%$  of which reflects our neglect of zero
point vibrations in the theory. The
lateral lattice parameter of the Cu substrate was set to the calculated
lattice constant for a fcc copper crystal. We chose a muffin tin (MT) radius
of $R^{\rm MT}_{\rm Cu}=2.20 \,\textrm{bohr}$ for the Cu atoms and a slightly
smaller radius $R^{\rm MT}_{\rm Co}=2.15\,\textrm{bohr}$ for the Co atoms to
prevent overlap of the MT spheres due to the strong relaxation found for some
systems.  

 The stability of various systems is analyzed with respect to the formation
 energy. Assuming that the slab is in thermal equilibrium with a Co and a Cu
 crystal, acting as reservoirs of Co and Cu atoms, the formation energy in eV
 per $(1\times 1)$-unit cell is defined as:
\begin{equation}
  \label{form_en}
  E^{\rm f}=\frac{1}{2A} \big(E^{\rm slab}-N_{\rm Cu}E_{\rm Cu}^{\rm bulk} 
      -N_{\rm Co}E_{\rm Co}^{\rm bulk}\big),
\end{equation}
where $A$ is the area of the surface unit cell of the considered
slab~\cite{area} and the factor $2$ accounts for the presence of two surfaces
of the slab. $N_{\rm Cu}$ and $N_{\rm Co}$ are the number of Cu and Co atoms in
the slab supercell and $E_{\rm Cu}^{\rm bulk}$ and $E_{\rm Co}^{\rm bulk}$ are
the energies of a Cu or a Co atom in the respective fcc bulk crystals at the
theoretical equilibrium lattice constants. Thus for a pure
Cu slab ($N_{\rm Co}=0$) $E^{\rm f}$ is the Cu surface energy, and for a pure
cobalt slab ($N_{\rm Cu}=0$) with $a_{\parallel}=a_{\rm Co}$ it is the surface
energy of cobalt.

The LAPW wave functions
 within the muffin tins (MTs) were expanded in spherical-harmonics with 
angular momenta up to $l_{\rm max}^{\rm wf}=10$. Non-spherical contributions 
to the electron density and potential within the MTs were considered up to
$l_{\rm max}^{\rm pot.}=4$. The cutoff for the Fourier-series expansion of the 
interstitial electron density and potential was chosen to be $G_{\rm max}=12.0\,
 \textrm{bohr}^{-1}$. 
\begin{table}
\begin{tabular}[h]{ c c c c}
$N_{{\bf k}_{\parallel}}$ & $E_{\rm cut}\,[\textrm{Ry}]$  &
$E^{\rm f}$~[eV/$(1\times 1)$-cell] & $\phi\,[\textrm{eV}]$ \\
\hline
6    & 15.6  & 1.50   & 5.27 \\
15   & 15.6  & 1.51   & 5.29 \\
21   & 15.6  & 1.51   & 5.29 \\
28   & 15.6  & 1.51   & 5.28 \\
36   & 15.6  & 1.51   & 5.28 \\
45   & 15.6  & 1.51   & 5.28 \\
21   & 12.8 &  1.58   & 5.31 \\
21   & 13.8  & 1.53   & 5.29  \\
21   & 17.5 &  1.50  & 5.27  \\
\end{tabular}
\centerline{
      \parbox{ 13.0cm}{
        \caption 
{\small{Convergence tests performed within LDA for a 5-layer slab of Co(001)
    strained at the lattice constant of copper and interlayer
    distance optimized for $N_{{\bf k}_{\parallel}}=28$. The surface energy
    $E^{\rm f}$ and work function $\phi$ are given as a function of the plane
    wave cut-off $E_{\rm cut}$ and the number of
    $\mathbf{k}_{\parallel}$-points in the irreducible part of the Brillouin
    zone $N_{{\bf k}_{\parallel}}$.}\label{tab:ecutnkpt}}}}
\end{table}
Extensive convergence tests with respect to $\mathbf{k}_{\parallel}$-point set
and the energy cutoff for the basis set were performed for a 5-layer Co(001)
slab at the lattice constant of copper and relaxed interlayer distance. The
results are shown in Table~\ref{tab:ecutnkpt}. A numerical accuracy  of $6 \%$
for the formation energy is achieved with $E_{\rm cut}=12.8\, \textrm{Ry}$,
while $E_{\rm cut}=15.6\, \textrm{Ry}$ is needed for an accuracy of $1
\%$. Thus a cutoff parameter of $15.6\, \textrm{Ry}$ was chosen throughout the
calculations. The Brillouin-zone integration was performed with a special
point set generated after the scheme of Monkhorst and Pack \cite{monkp}. We
obtained an accuracy of the Brillouin-zone integration better than $1\%$
by using 21 $\mathbf{k}_{\parallel}$-points in the irreducible wedge of the
Brillouin-zone (IBZ) (see Table \ref{tab:ecutnkpt}). 

The bulk  energies needed as a reference to determine the formation
energy (see Eq.(\ref{form_en})) were calculated using
the same LAPW-parameters as in the slab calculations. For the bulk calculation
104 $\mathbf{k}$-points in the IBZ were used.

\begin{table}
\begin{tabular}[h]{ c c c c }
$N_{\rm layer}$ & $E^{\rm f}$~[eV/$(1\times 1)$-cell]  & $\phi\,[\textrm{eV}]$
& $\Delta d_{\rm 12}/d_0 [\%]$\\ 
\hline
3    & 0.79 & 4.91 & -2.93 \\
5    & 0.78 & 4.78 & -3.01 \\
7    & 0.78 & 4.83 & -3.10 \\
9    & 0.78 & 4.82 & -3.11 \\
\end{tabular}
\centerline{
      \parbox{ 13.0cm}{
        \caption 
{\small{DFT-LDA results for the surface
    energy $E^{\rm f}$, work functions $\phi$, and relaxation $\Delta d_{\rm
      12}/d_0$ of the clean Cu(001)-surface are given as a function of the
    number of slab layers $N_{\rm layer}$, $N_{\bf k}=21$, $E_{\rm cut}=15.0$
    [Ry].  }\label{tab:cuNlayer}}}}
\end{table}
Prior to investigating the effects of adsorption of cobalt, we checked the
required thickness of a copper slab, to ensure a good representation of the
properties of the clean Cu(001)-surface. The surface energies, work functions
and interlayer relaxations for slabs of 3, 5, 7, and 9 layers of copper are
compared in Table \ref{tab:cuNlayer}. For a 5-layer slab the relaxation between
surface and subsurface layer of $-3.01\%$ is close to the experimental
 value  obtained by MEIS\cite{jiang} ($-2.4\%$) while the
 LEED-result\cite{davno} is smaller ($-1.1\pm0.4\%$). The calculated work
 function $\phi = 4.78\, \textrm{eV}$ is in good agreement with experiment:
 $4.59\pm0.05\, \textrm{eV}$ \cite{workfunc1}, $4.76\, \textrm{eV}$
 \cite{workfunc2}, and $4.77\pm0.05\, \textrm{eV}$ \cite{workfunc3}.
\begin{table}
\begin{tabular}[h]{ c c c   }
$N_{\rm layer}$ & $E^{\rm f}$~[eV/$(1\times 1)$-cell]  & $\phi
\,[\textrm{eV}]$   \\ 
\hline
3    & 1.767 & 5.31         \\
5    & 1.754 & 5.31         \\
7    & 1.753 & 5.30         \\
9    & 1.759 & 5.30         \\
\end{tabular}
\centerline{
      \parbox{ 13.0cm}{
        \caption 
{\small{Formation energy and work functions for 1Co/Cu(001) with optimized
    interlayer distances as a
    function of the number of substrate layers $N_{\rm layer}$.
      }\label{tab:cosubthick }}}}
\end{table}

A further  requirement of the thickness of the slab is that the
interaction of the Co layers on both sides of the slab trough
the substrate is negligible for the questions of concern. To test the strength
of this interaction we studied the formation energy and work function 
for 1ML Co on Cu(001) (1Co/Cu(001)) as a function of the substrate
thickness. The results are summarized in Table \ref{tab:cosubthick }.
Both the formation energy and work function converge quickly with the
substrate thickness. On the basis of these results we conclude that a 5-layer
copper slab represents a good approximation of the Cu(001)-surface.

\section{Formation Energy and Stability}\label{stabil}
\begin{figure}
\centering
\centerline{\psfig{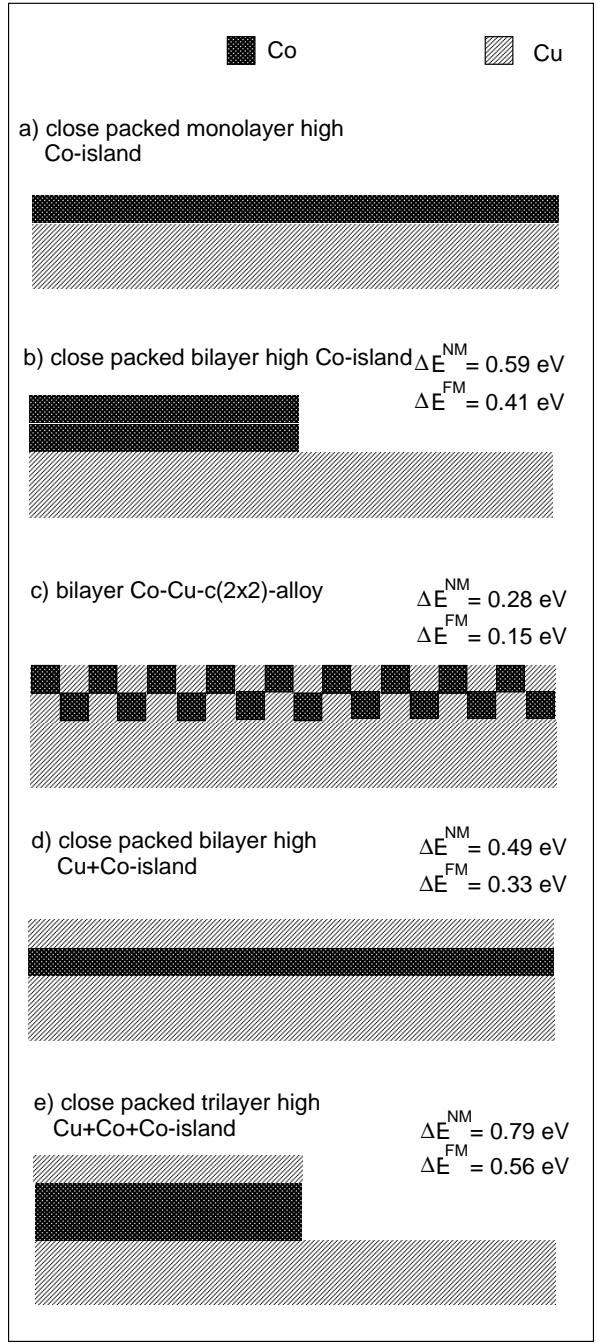}}
      \vspace*{0.3cm}
   \caption{\label{fig:conf1.0} Schematic diagram of the different
     adsorbate configurations for a Co coverage of $\Theta=1.0$ ML. The
     formation energy changes for configurations b)-e) in the
     nonmagnetic $\Delta E^{\rm NM}$ and ferromagnetic case $\Delta E^{\rm
       FM}$ is given with respect to formation energy of the close packed
     monolayer high island shown in a).}
\end{figure}

In order to identify the equilibrium configuration of Co on Cu(001) we
investigate in this Section the tendency towards separation in multilayer
islands and the influence of a copper capping layer. The studied systems include
the clean Cu(001) surface, a monolayer and bilayer thick cobalt film on
Cu(001) denoted by 1Co/Cu(001) and 2Co/Cu(001), respectively, as well as the
corresponding capped systems, 1Cu/1Co/Cu(001) and
1Cu/2Co/Cu(001). Additionally we investigated a bilayer Co-Cu-c$(2\times
2)$-alloy. The calculated formation energies and the work functions are given
in Table~\ref{formen_tab}. 
\begin{table}
\begin{tabular}[h]{ c c c }
System   & $ E^{\rm f}$~[eV/$(1\times 1)$-cell] & $\phi$~[eV]      \\
\hline
Cu(001)   & 0.78     & 4.78    \\
\hline
1Co/Cu(001) NM  & 1.75     & 5.31    \\
1Co/Cu(001) FM & 1.51      & 5.16    \\
\hline
2Co/Cu(001 NM  &   1.55    & 5.38      \\
2Co/Cu(001 FM  &  1.42     & 4.88  \\
\hline
1Cu/1Co/Cu(001) NM &   1.27           & 4.89      \\
1Cu/1Co/Cu(001) FM &   1.18           & 4.81       \\
\hline
1Cu/2Co/Cu(001) NM &  1.14            & 4.74      \\
1Cu/2Co/Cu(001) FM &  1.12            & 4.82  \\
\hline
bilayer Co-Cu-c$(2\times 2)$-alloy NM  & 1.48     & 5.18    \\
bilayer Co-Cu-c$(2\times 2)$-alloy FM & 1.36      & 4.97    \\
\end{tabular}
\centerline{
      \parbox{ 13.0cm}{
        \caption 
{\small{Formation energies $E^{\rm f}$ and the work functions $\phi$ for
    various structures.}\label{formen_tab}}}}
\end{table}
We consider several ways in which a total coverage of 1ML Co can be arranged
on a Cu(001) surface. The energy of a system consisting of more than one
domain, namely regions of clean copper surface  and regions covered 
by large cobalt islands, is simply given by the weighted sum of the
formation energy of the clean Cu(001) surface and those of the
Co-island. Under the assumption that the islands are large, the contributions
of the step edges and side facets of the islands are negligible and were not
taken into account. A schematic presentation of the different structures are
given in Fig. \ref{fig:conf1.0} together with energy changes with respect to
the case where the whole surface is covered by a monolayer thick $(1\times
1)$-cobalt layer.  

Our calculations show that a monolayer film, Fig.~\ref{fig:conf1.0}(a),
would separate into a clean Cu(001)-surface and a bilayer
island, Fig.~\ref{fig:conf1.0}(b). For the nonmagnetic case the gain in energy
is $\Delta E^{\rm NM} = 0.59$ eV/$(1\times 1)$-cell and for the ferromagnetic
case it is $\Delta E^{\rm FM} = 0.41$ eV/$(1\times 1)$-cell. This result can
be explained in terms of the higher coordination of the cobalt atoms in the
bilayer film and correlates with the substantial broadening of the cobalt
$d$-band and the strong relaxation between Co layers in 2Co/Cu(001) as will be
discussed later in this paper. Concerning the effect of magnetism, we see that
it reduces, but does not change the tendency towards formation of bilayer
islands.

Experimental studies\cite{lit,kir2} show that copper segregates onto the
surface after annealing. Therefore we study here the influence of a copper
capping layer on stability. Covering 1Co/Cu(001) with a monolayer of copper,
Fig.~\ref{fig:conf1.0}(d), reduces the energy of the system by $\Delta E^{\rm
  NM} =0.49$~eV/$(1\times 1)$-cell. Compared to the cobalt terminated systems
the copper-capped systems gain less spin-polarization energy because of the
hybridization with the capping layer. Consequently the energy gain due to a
capping layer for the magnetically ordered system is lower than the one for
the nonmagnetic: ($\Delta E^{\rm FM}=0.33$~eV/$(1\times 1)$-cell). The
influence of the capping layer on the magnetic properties of the copper
covered systems will be discussed in Section~\ref{magn}. Sa{\'u}l
and Weissmann~\cite{mweiss} recently calculated the surface segregation energy
of $3d$-impurities (Fe, Co, Ni) in Pd, Ag, and Cu. They found, in agreement
with our results, that embedding in the bulk of the host material is
connected with a substantial gain in energy both for
nonmagnetic and magnetic impurities/(layers), the effect being weaker for the
latter. We also note that the
copper capping layer in the 1Cu/1Co/Cu(001) and 1Cu/2Co/Cu(001) systems has
properties similar to the clean Cu(001)-surface; for example we find that the
work functions of the systems are $\phi_{\rm Cu(001)} = 4.78\, \textrm{eV}$,
$\phi_{\rm 1Cu/1Co/Cu(001)} = 4.89\,  \textrm{eV}$, and $\phi_{\rm
  1Cu/2Co/Cu(001)} = 4.74\, \textrm{eV}$.   

Recent combined STM and RHEED experiments~\cite{nouv} detected ordered
c$(2\times 2)$ regions when a total coverage of 1ML Co was deposited on
Cu(001) at room temperature and subsequently annealed at $450$~K. Motivated by
these results, we studied a configuration, where starting from 1Co/Cu(001)
every other Co atom is replaced by a Cu atom in the substrate layer
underneath. In this way a bilayer $c(2 \times 2)$-alloy~\cite{bil-c2x2}, shown
schematically in Fig.~\ref{fig:conf1.0}(c), is formed. We find that this
configuration is by $0.28$~eV/$(1\times 1)$-cell (nonmagnetic) and
$0.15$~eV/$(1\times 1)$-cell (ferromagnetic case) more favorable than the
$(1\times 1)$-monolayer in Fig.~\ref{fig:conf1.0}(a). However, it is a
metastable structure because transition into a cobalt monolayer covered by
copper, 1Cu/1Co/Cu(001) in Fig.~\ref{fig:conf1.0}(d), leads to an energy gain of
$0.22$~[eV/$(1\times 1)$-cell] (nonmagnetic) and $0.18$~eV/$(1\times 1)$-cell
(ferromagnetic case). Thus, the bilayer $c(2 \times 2)$-alloy lies
energetically between the 1Co/Cu(001) and 1Cu/1Co/Cu(001) systems
and it may be stabilized kinetically. This surface alloy might also
represent a favorable configuration with respect to surface strain
relief. Indeed a $c(2 \times 2)$-pattern was observed experimentally
preferentially in the middle of large islands~\cite{poels}.  

We also studied whether the cobalt $(1\times 1)$-layer will prefer to be
buried deeper in the substrate. Our calculations for the 1Cu/1Co/Cu(001) and
2Cu/1Co/Cu(001) systems show that there is no additional energy gain through
covering the system with a thicker copper layer.

\begin{table}   
\begin{tabular}[h]{ c c c   }
System & $E^{\rm I}_{\rm NM}$  & $E^{\rm I}_{\rm FM}$   \\
\hline
\ldots Cu/1Co/Cu\ldots    & 0.163 & 0.113        \\
\ldots Cu/2Co/Cu\ldots    & 0.044 & 0.066        \\
\ldots Cu/3Co/Cu\ldots    &-0.005 & 0.046        \\
\end{tabular}
\centerline{
      \parbox{ 13.0cm}{
        \caption 
{\small{Interface energies given in [eV/$(1\times 1)$-cell] for nonmagnetic
    ($E^{\rm I}_{\rm NM}$)and ferromagnetic ($E^{\rm I}_{\rm FM}$) systems
    as a function of the cobalt interlayer thickness.
      }\label{tab:e_inter}}}}
\end{table}
The segregation of Cu on the surface is typically explained by the lower
surface energy of Cu(001) compared to Co(001). Still this argument is only
applicable if the interface energy were small and thus negligible. In order to
calculate the energy cost to create an interface we studied three different
systems with one, two and three cobalt interlayers in copper bulk, marked as
(\ldots Cu/1Co/Cu\ldots), (\ldots Cu/2Co/Cu\ldots) and  (\ldots
Cu/3Co/Cu\ldots), respectively. Here the lateral parameter is set to the
copper bulk value while the interlayer distances are relaxed. The interface
energies, $E^{\rm I}$, are calculated analogously to the formations energies
[compare Eq.(\ref{form_en})]. In order to subtract the effect of elastic
strain due to the lattice mismatch of the two materials we use as a reference
energy for cobalt the bulk energy of fct cobalt with $a_{\parallel}=a_{\rm
  Cu}$ and relaxed $a_z$~\cite{co_fct} instead of the one of fcc cobalt at the
cobalt lattice constant. Table~\ref{tab:e_inter} lists the results for
nonmagnetic and ferromagnetic cobalt interlayers. It shows that the values of
$E^{\rm I}$ for a cobalt bilayer and trilayer are very close, i.e. the
interface energy converges quickly with the thickness of the cobalt layer.  
The interface energy is indeed significantly smaller than the difference of
the surface energies of the clean Cu(001)-surface ($E^{\rm f} =
0.79$~eV/$(1\times 1)$-cell) and a thick Co(001) film with a lateral parameter
fixed to the lattice constant of copper and relaxed interlayer distances
($E^{\rm f}_{\rm NM} = 1.21$~eV/$(1\times 1)$-cell~\cite{e_co_fct} and $E^{\rm
  f}_{\rm FM} = 1.11$~eV/$(1\times 1)$-cell~\cite{e_co_fct}). For this reason
the common argument that simply the surface energy difference of cobalt and
copper explains the segregation of substrate material on the surface works in
the case of Co on Cu(001).  

In analogy to the Co terminated system, the single Co layer capped
by Cu shown in Fig.~\ref{fig:conf1.0}(d) will tend to separate into a clean
Cu(001) surface and a double Co layer capped by Cu,
Fig.~\ref{fig:conf1.0}(e). Still the energy gain due to phase separation
[$\Delta E^{\rm NM} = 0.31$~eV/$(1\times 1)$-cell, $\Delta E^{\rm FM} =
0.23$~eV/$(1\times 1)$-cell)] is only about half the energy
gain for the system with Co on the surface. We can summarize that both the
magnetic ordering and the capping layer weaken the tendency towards cobalt
clustering but {\em qualitatively} we observe the same behavior with and
without magnetism. 
\begin{figure}
\centering
\centerline{\psfig{figure=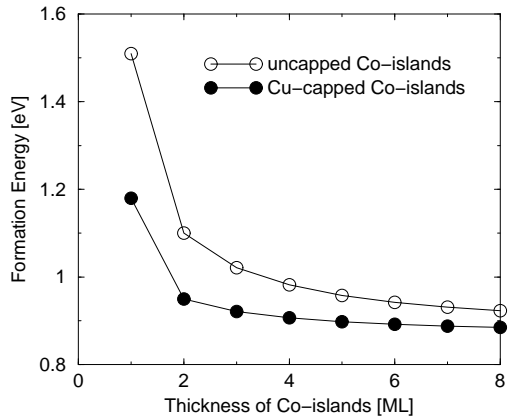,width=6.6cm,angle=270}}
      \vspace*{0.3cm}
   \caption{\label{e_sep.eps} Formation energy of
     different ferromagnetically ordered configurations for a total cobalt coverage
     of 1~ML as a function of the
     cobalt island thickness $N$. The structures consist of clean Cu(001) and
     a compact island with $N$ Co layers ($\circ$) or $N$ Co-layers capped by
     copper. The area covered by the cobalt islands is $\frac{1}{N}$ of the
     whole surface. Especially for the copper terminated systems the
     separation in higher than bilayer cobalt islands is unlikely because of a
     negligible energy gain. }
\end{figure}

Yet we need to find out whether the bilayer film will be stable or if
higher cobalt islands may form. In order to determine the formation energy of a
$N$-layer ($N>2$) thick cobalt island we assume that each of the {\em
  intermediate} cobalt layers has an energy of a bulk atom in a tetragonal
cobalt crystal with $a_{\parallel}=a_{\rm Cu}$ and fully relaxed
$a_{z}$~\cite{co_fct}. The elastic energy contribution is the difference
between the energy of cobalt bulk at the fcc cobalt lattice constant and the
energy of fct cobalt as described above and amounts to $0.11$~eV (NM) and
$0.08$~eV (FM) per cobalt atom. Thus the formation energy of a $N$-layer
cobalt film is 
\begin{equation}
  \label{form_en-N}
  E^{\rm f}_{N{\rm Co/Cu(001)}}=E^{\rm f}_{\rm 2Co/Cu(001)}+(N-2)E^{\rm elast.}.
\end{equation}
For a total coverage of 1ML the formation energy of a configuration consisting
of clean Cu(001) surface and a $N$-layers high cobalt island is given by:
\begin{equation}
  \label{form_en_Nisl}
  E^{\rm f}_{N{\rm Co-island}}=\frac{1}{N}E^{\rm f}_{N{\rm
    Co/Cu(001)}}+\frac{N-1}{N}E^{\rm f}_{\rm Cu(001)}.
\end{equation}
An analogous expression for the formation energy holds for the copper capped
systems. The formation energy of the ferromagnetically ordered capped and
uncapped systems is plotted in  Fig.~\ref{e_sep.eps} as a function of
cobalt-island height $N$. In the following we will concentrate on the
Cu-terminated systems because, as can be seen from Fig.~\ref{e_sep.eps}, they
are always lower in energy. The substantial energy gain due to separation of a
monolayer-thick cobalt film in bilayer islands was already
discussed above. Yet further separation in higher cobalt
islands brings only a small energy gain, e.g., the gain due to separation
from bilayer in trilayer islands for the Cu-terminated islands is
about $0.03$~eV/$(1\times 1)$-cell. We note that this
energy gain is mainly due to the increase of the clean Cu(001)-surface. 
Moreover the cost of the island facets, which was not taken
into account in the present discussion, grows with island height. Because of
the small energy gain, as shown in Fig.~\ref{e_sep.eps}, and the
increasing cost of the sidewalls the formation of islands higher than bilayer
is unlikely. We conclude that the ferromagnetically ordered configuration in
  Fig.~\ref{fig:conf1.0}(e), which is by $0.56$~eV/$(1\times 1)$-cell more
  favorable than the one in Fig.~\ref{fig:conf1.0}(a), represents the
  thermodynamic equilibrium structure.

\section{Structural Properties}\label{struct}
In the previous Section we identified the bilayer cobalt island covered by a
copper capping layer as the thermodynamically stable structure. However,
crystal growth represents a situation which is more or less far from
thermodynamic equilibrium therefore not only the equilibrium structure but
also other, metastable, structures (e.g. those
shown in Fig.~\ref{fig:conf1.0}) may occur. In this Section we 
present the results of a 
geometry optimization for monolayer and bilayer
$(1\times 1)$- as well as copper-capped systems. The Co films are assumed to
grow pseudomorphically on the Cu(001)-surface, adopting the lateral spacing of the
Cu crystal. The calculated relaxation of the interlayer spacing for the
nonmagnetic and ferromagnetic systems is given in Table~\ref{struct_tab}. We
remind the reader that all relaxations are given with respect to the interlayer 
spacing in copper bulk.
\begin{table}
\begin{tabular}{ c c c  }
Method   & $\Delta d_{\rm 12}/d_0$ & $\Delta d_{\rm 23}/d_0$     \\
         & $[\%]$       & $[\%]$       \\
\hline
\multicolumn{3}{c}{1ML Co/Cu(001)}  \\
\hline
FP-LAPW NM       & $-4.7\%$ &  $-0.3\%$    \\
FP-LAPW FM       & $-3.0\%$   & $0.0\%$     \\
LEED\cite{clarke} &$-6.0\%$     & $-6.0\%$  \\
LEED\cite{kir1} & $-2.5\%$ & $-1.4\%$   \\
\hline
\multicolumn{3}{c}{2ML Co/Cu(001)}  \\
\hline
FP-LAPW NM     & $-17.0\%$    & $0.0\%$   \\
FP-LAPW FM     & $-13.4\%$  & $-0.8\%$   \\
LEED\cite{kir1} & $-2.0\%$   & $-4.2\%$    \\
\hline
\multicolumn{3}{c}{1ML Cu/1ML Co/Cu(001)}  \\
\hline
FP-LAPW NM     & $-7.0\%$    & $-3.0\%$  \\
FP-LAPW FM     & $-5.6\%$    & $-2.0\%$  \\
\hline
\multicolumn{3}{c}{1ML Cu/2ML Co/Cu(001)}  \\
\hline
FP-LAPW NM     & $-5.0\%$    & $-14.8\%$  \\
FP-LAPW FM     & $-4.6\%$    & $-12.5\%$   \\
\end{tabular}
\centerline{
      \parbox{ 13.0cm}{
        \caption 
{\small{Relaxation $\Delta d_{\rm 12}/d_0$ and $\Delta d_{\rm 23}/d_0$ of the
    interlayer spacing in $\%$ for the first two layers compared to the
    lattice parameter of Cu bulk, $d_0$.}\label{struct_tab}}}} 
\end{table}

The first interlayer spacing in the monolayer film, $d_{\rm Co-Cu}$,
shows an inward relaxation of $4.7\%$ for the nonmagnetic case, which reduces
to $3.0\%$ for the ferromagnetic film. At the same time the interlayer spacing 
between the first and second substrate layer, which is contracted by $3.0\%$
for the clean Cu(001) surface, expands back to the bulk value
$-0.3\%$ ($0.0\%$) for the nonmagnetic (ferromagnetic) system. This result
can be explained in terms of the bond-cutting model. Due to the missing bonds
of the surface atoms the strength of the remaining bonds to the subsurface
layer is enhanced, giving rise to an inward relaxation~\cite{methf1}. 
The bond strength is also related to the $d$-band
occupation~\cite{methf1}, thus Co-Cu bonds are stronger than Cu-Cu bonds and
consequently upon cobalt adsorption we observe a stronger relaxation 
of the Co-Cu-interlayer distance, while the Cu-Cu distance expands.
Previous {\em ab initio} results\cite{wufre} found a relaxation of the first
interlayer distance of the ferromagnetic 1Co/Cu(001) surface of $-10.4
\%$. The reason for the discrepancy with our result (which is $-3.0\%$) 
is in the choice of the lateral lattice parameter.
While we use a non-relativistic treatment of the valence electrons 
and the corresponding theoretical equilibrium lattice constant of Cu
(3.55~{\AA}), Wu and Freeman\cite{wufre} used a semi-relativistic 
treatment which gives a noticeably smaller lattice constant
(3.52~{\AA}). Nevertheless, in their adsorbate study Wu and Freeman\cite{wufre} 
set the lateral lattice parameter to the substantially larger 
experimental  value (3.61~{\AA}). As a consequence,
the strong interlayer relaxation found by Wu and
Freeman\cite{wufre} just reflects that their copper surface is
under tensile strain.  We tested this and indeed could reproduce
the effect: When we use a semi-relativistic treatment of the valence electrons
and still force the Cu substrate to assume the experimental lattice
constant we obtain $\Delta d_{\rm 12}/d_0=-7.9\%$ and $\Delta d_{\rm
  23}/d_0=-2.3\%$.

For the bilayer Co film we obtain a surprisingly strong contraction of the
interlayer distance of $d_{\rm Co-Co}$ of $-17\%$ in the nonmagnetic
case. For the ferromagnetically ordered system the contraction is somewhat
smaller, $-13.4\%$, due to the magnetovolume effect. These results can hardly
be explained by comparing the lattice constant of the fcc bulk phase
of Co with that of bulk Cu. Such comparison would give a lattice mismatch  of
$-4.3 \%$ in the nonmagnetic and $-2.8 \%$ in the ferromagnetic case. 
Thus, in such description one would say that the Co film is strained, but that
the effect is not very large. However, it is questionable whether the
comparison of the bulk lattice parameters of the two materials, is a good
approach for understanding ultrathin films with $\Theta \le 2$ ML. For
example, total-energy calculations based on the FP-LAPW method~\cite{dipl96}
show that the difference between the equilibrium lattice constant of a {\em
  free standing} cobalt monolayer and that of the Cu substrate is $-14.1 \%$
for a nonmagnetic and $-12.2 \%$  for a ferromagnetic
monolayer~\cite{umls}. Therefore, if we refer the strain in the Co adlayers to 
the lattice parameter of the free-standing Co layer, the strain is
significant, and the above noted interlayer relaxation then simply reflects
the reaction of the Co film to this big strain. Indeed, we think that this
description is appropriate (in a qualitative sense) because for a very thin
cobalt film ($\Theta \leq 2$~ML) the bonding to the noble metal substrate can
only partially replace the bonds to missing cobalt neighbors. Thus, the
adsorbed film will still bear some resemblance to the free-standing one. The
above result also indicates that the weaker binding to the substrate is balanced by
forming a strong bond between the two cobalt layers.

The competition between Co-Co and Co-Cu-bonding is also a driving force for the
structural changes in the capped systems. The hybridization with the
copper capping layer has the general effect of weakening the existing Co-Cu and
Co-Co bonds in the 1Co/Cu(001) and the 2Co/Cu(001) system,
respectively. Consequently, the
interlayer distance between the cobalt and copper layer increases from
$-4.7\%$ (NM) and $-3.0\%$ (FM)  in 1Co/Cu(001) to $-3.0\%$ (NM) and $-2.0\%$
(FM)  in 1Cu/1Co/Cu(001). Similarly the strong relaxation between the two
cobalt layers decreases from $-17.0\%$ (NM) and $-13.4\%$ (FM) in 2Co/Cu(001)
to $-14.8\%$ (NM) and $-12.5\%$ (FM)  in 1Cu/2Co/Cu(001). On the other hand
the stronger Co-Co bond induces a weaker binding with the
capping layer which is reflected in the smaller inward relaxation of the
distance between the capping layer and the cobalt film of $-5.0\,\%$
for the 1Cu/2Co/Cu(001) system compared to $-7.0\,\%$ for the 
1Cu/1Co/Cu(001) system.

Table~\ref{struct_tab} also contains structural
data determined with LEED. We note, however, that such structural
analysis is complicated and not unambiguous, because, as discussed above,
in the Co/Cu(001) system several domains and/or metastable structures 
may coexist. In the absence of knowledge about these various structures
and their energies, it appeared  to be a reasonable choice
for Clarke {\em et al.}~\cite{clarke} to assume that Co on Cu(001) will
form a full Co monolayer. And based on this assumption they
determined an inward relaxation of $-6\%$ for both $d_{\rm Co-Cu}$ 
and $d_{\rm Cu-Cu}$. Our work, however, shows that the 1Co/Cu(001) 
system is unstable with respect to the formation of bilayer islands 
and capped bilayers structures.  In a more recent LEED study 
Cerda {\em et al.}~\cite{kir1} assumed the coexistence of regions of clean
Cu(001), Co monolayer and Co bilayer islands, but Co layers
with a Cu capping layer, which we find to have the lowest total
energy, were not considered. Thus, in both experimental analyses
the model assumptions didn't include all relevant systems.

\section{Magnetic Properties}\label{magn}
The layer-resolved magnetic moments in the four systems studied, 1Co/Cu(001),
2Co/Cu(001), 1Cu/1Co/Cu(001) and 1Cu/2Co/Cu(001) are given in
Table~\ref{magnm_tab}. To be precise, these are the contributions from the
muffin tin region only. 
\begin{table}[tb!]
\begin{tabular}{ |c|c|c| }
System          &  layer       & $M~[\mu_{\rm B}]$ \\ 
\hline
1Co/Cu(001)     & Cu(C)     & -0.004 \\
                & Cu(S-2)   & -0.014 \\
                & Cu(S-1)   & 0.024  \\
                & Co(S)     & 1.711 \\
\hline
2Co/Cu(001)     & Cu(C)     & -0.002 \\
                & Cu(S-3)   & -0.009 \\
                & Cu(S-2)   & 0.016  \\
                & Co(S-1)   & 1.472 \\
                & Co(S)     & 1.706 \\
\hline
Co(001) at $a_{\rm Cu}$     & Co(C)     & 1.648  \\
                             & Co(S-1)   & 1.615 \\
                             & Co(S)     & 1.783 \\
\hline
1Cu/1Co/Cu(001) & Cu(C)     & -0.001 \\
                & Cu(S-3)   & -0.008 \\
                & Cu(S-2)   & 0.027  \\
                & Co(S-1)   & 1.445 \\
                & Cu(S)     & 0.040 \\
\hline
1Cu/2Co/Cu(001) & Cu(C)     & -0.001 \\
                & Cu(S-4)   & -0.007 \\
                & Cu(S-3)   & 0.022  \\
                & Co(S-2)   & 1.383 \\
                & Co(S-1)   & 1.374 \\
                & Cu(S)     & 0.035 \\
\end{tabular}
\centerline{
      \parbox{ 13.0cm}{
        \caption 
{\small{Layer resolved local magnetic moments in the ferromagnetic systems in
    $[\mu_B]$ as obtained from the slab calculation. S, S-1, S-2, etc. denote
    the position of the corresponding layer with respect to the surface, S
    being the surface layer and C being the central layer of the slab.
      }\label{magnm_tab}}}}
\end{table}
The top layer in 1Co/Cu(001) exhibits an
enhanced magnetic moment ($M_{\rm Co(S)}=1.71\,\mu_B$) compared to the bulk
value of $1.52\,\mu_B$, calculated at the equilibrium lattice constant of
cobalt. This is due to the larger lateral constant of the epitaxial cobalt
adlayer and to the reduced coordination on the surface. 
Further we find that the surface layer of the 2Co/Cu(001) system
exhibits the same magnetic moment ($1.71\,\mu_B$) as the 1Co/Cu(001)
system. In fact, a thick fcc cobalt film at the lattice constant of 
copper also has a similar moment, namely $1.78\,\mu_B$. 
However, the magnetic moment of the subsurface Co layer,
which binds to the Cu substrate, is reduced 
to $1.47\,\mu_B$. The corresponding magnetic
moment of subsurface cobalt in a thick fcc cobalt film at
the lattice constant of copper is $1.62\,\mu_B$. The lower magnetic moment of
subsurface cobalt is a consequence of the higher coordination and the strong
contraction of the interlayer spacing $d_{\rm Co-Co}$.
 
The hybridization with the copper capping layer reduces the magnetic moment of
the first Co layer both in the 1Cu/1Co/Cu(001) and 1Cu/2Co/Cu(001) systems
by about $0.3\,\mu_B$ compared to the 1Co/Cu(001) and 2Co/Cu(001) systems.
It is interesting to note that both Co layers in 1Cu/2Co/Cu(001) have the
same magnetic moment ($1.38\,\mu_B$) which can be explained by the fact 
that Co(S-1) and Co(S-2) have the same coordination of Co and Cu atoms.

To our knowledge  magnetic moments for 1Co/Cu(001) have not yet been 
measured due to the already discussed difficulties in the preparation
of a single cobalt monolayer on Cu(001). Our calculated  value
of the surface magnetic moment of 1Co/Cu(001), $1.71\,\mu_B$, is 
slightly lower than that obtained in previous calculations,
e.g. $1.78\,\mu_B$~\cite{wufre} and $1.76\,\mu_B$
\cite{sb2} from FP-LAPW
and $1.85\,\mu_B$~\cite{lmto2} from FP-LMTO
(full-potential linearized muffin tin orbitals)
calculations. The differences are attributed mainly to the use of the
experimental lattice constant of copper (3.61~{\AA}) and/or the lack of
considering the interlayer relaxation in Refs.\cite{wufre,sb2,lmto2} (see also
Section~\ref{struct}).

The magnetic moments of 1.9 ML and 2.1 ML of Co deposited on Cu(001) 
measured with x-ray magnetic circular dichroism (XMCD) are
$1.71\pm0.1\,\mu_B$\cite{babs2} and $1.77\pm0.1\,\mu_{\rm B}$\cite{babs1},
respectively. The XMCD-spectra were recorded at $40$~K, but 
information about the preparation conditions, which could 
tell whether the Co layers were capped by Cu, is not available.
Yet, the magnetic moment compares well with our calculated magnetic 
moment for 2Co/Cu(001). 
With respect to other theoretical work, we note that
the same trend of an enhanced magnetic moment in the
surface layer ($1.85\,\mu_B$) and a reduced  magnetic moment in the subsurface
layer ($1.60\,\mu_B$) was found in a previous FP-LAPW calculation for
2Co/Cu(001)~\cite{sb3}. In this study the lateral parameter was fixed 
to the experimental lattice constant of copper and relaxation
of the interlayer spacing was not taken into account.
However, the strong relaxation of the interlayer spacing in
2Co/Cu(001) discussed in  Section~\ref{struct} has a noticeable influence on the
magnetic moments and cannot be neglected.

Our results reveal also that the adsorbed cobalt film
induces a small polarization in the substrate. The magnetic moment of
the copper layer at the interface is positive, e.g. in 1Co/Cu(001) 
it is $0.024\,\mu_B$. Then, in the next layer it switches to
a negative value ($-0.014\,\mu_B$). Also the central 
layer of our 5-layer Cu slab has a very small negative 
moment, $-0.004\,\mu_B$. The oscillation of the magnetic moment 
perpendicular to the surface indicates the formation of a spin-density 
wave. This striking effect is observed for all
four studied systems. However, we note that for a detailed investigation of
this effect, a thicker substrate slab has to be considered. The magnetic
moment induced in the capping layer is somewhat larger than the one induced in
the substrate layer: $0.040\,\mu_B$ in 1Cu/1Co/Cu(001) and $0.035\,\mu_B$ in
1Cu/2Co/Cu(001).

\section{Electronic Properties}\label{electr}
\begin{figure}
\centering
\centerline{\psfig{figure=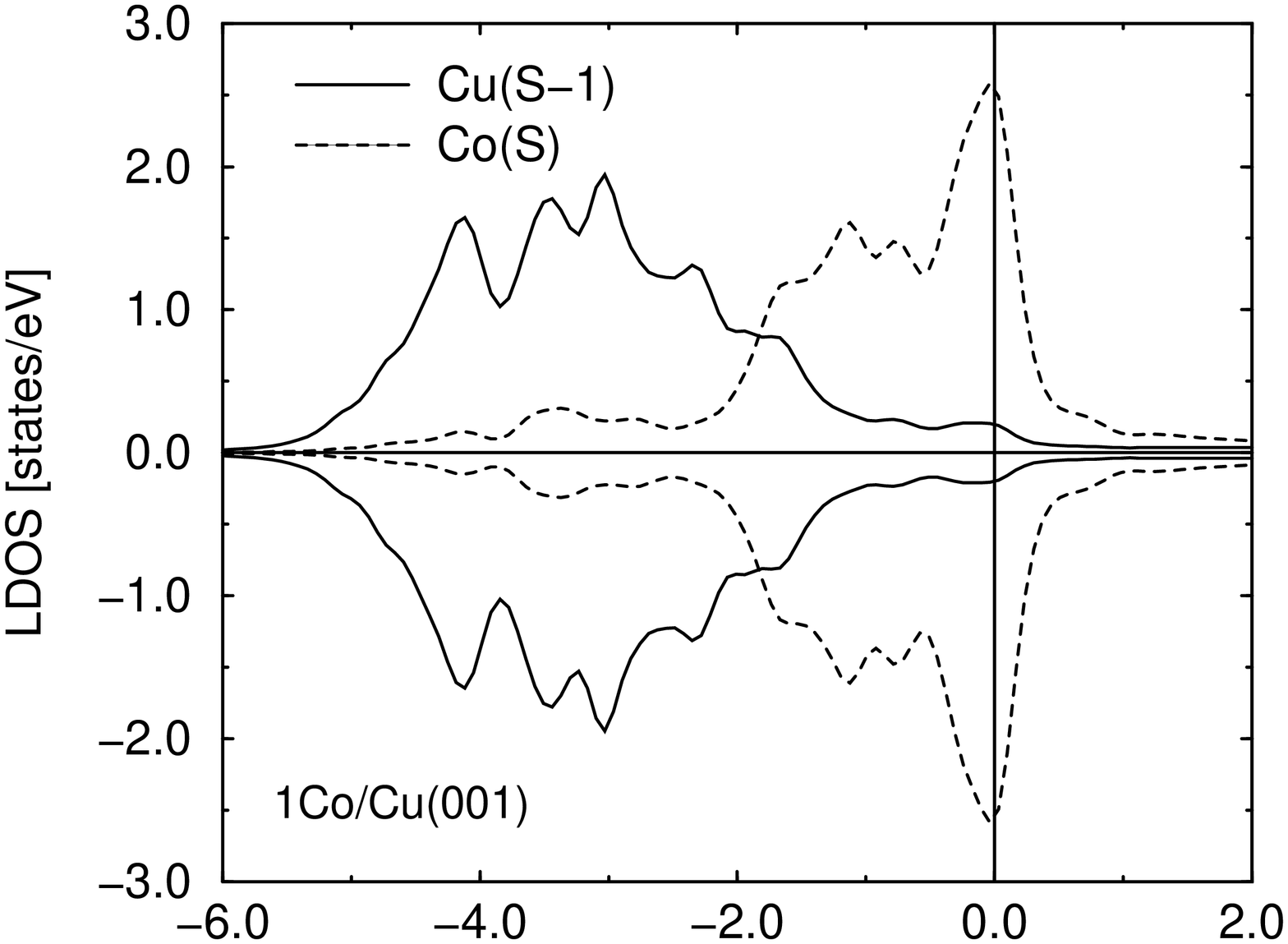,width=6.5cm,angle=270}}
      \vspace*{-0.8cm}
\centerline{\psfig{figure=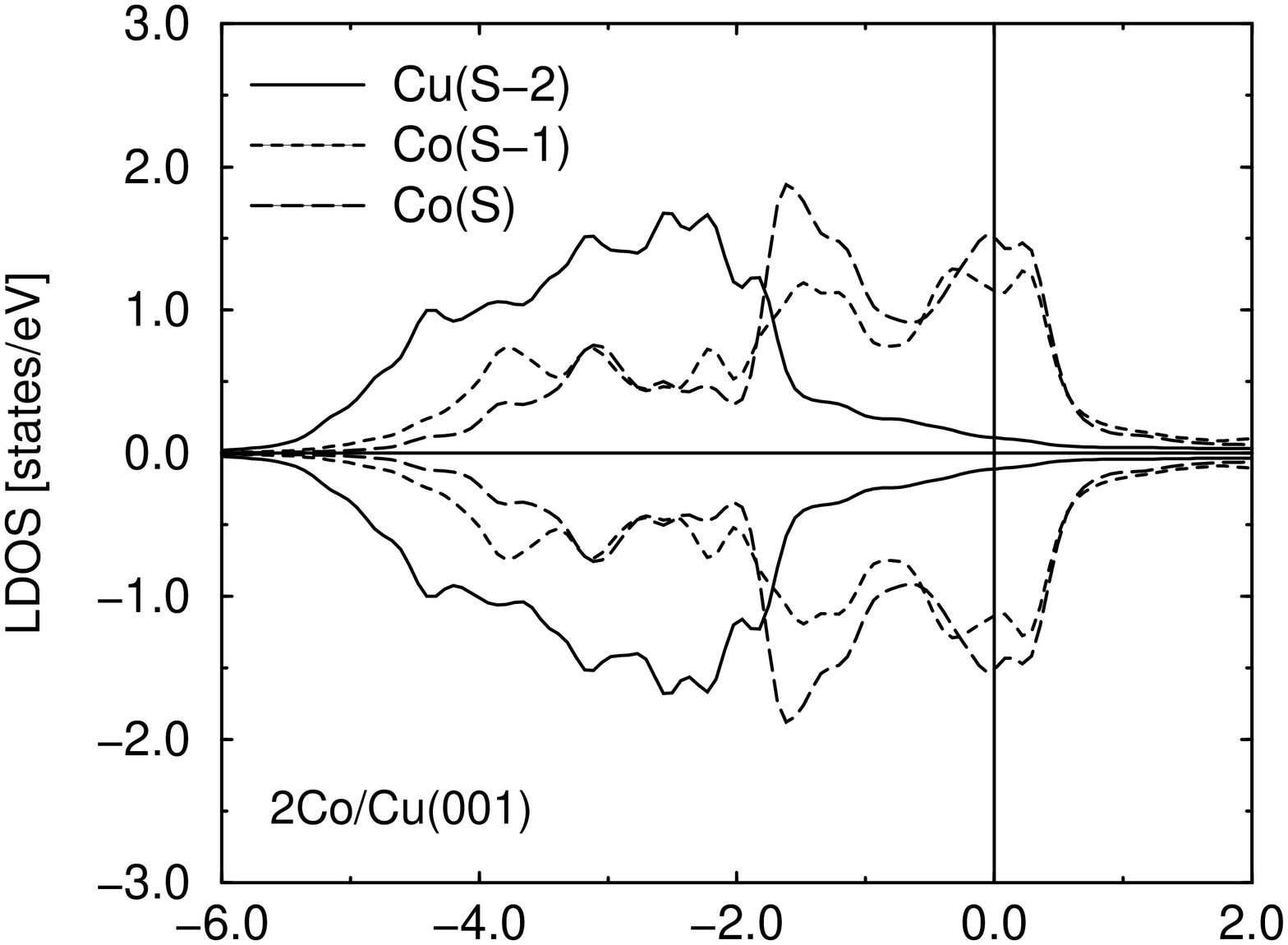,width=6.5cm,angle=270}}
      \vspace*{-0.8cm}
\centerline{\psfig{figure=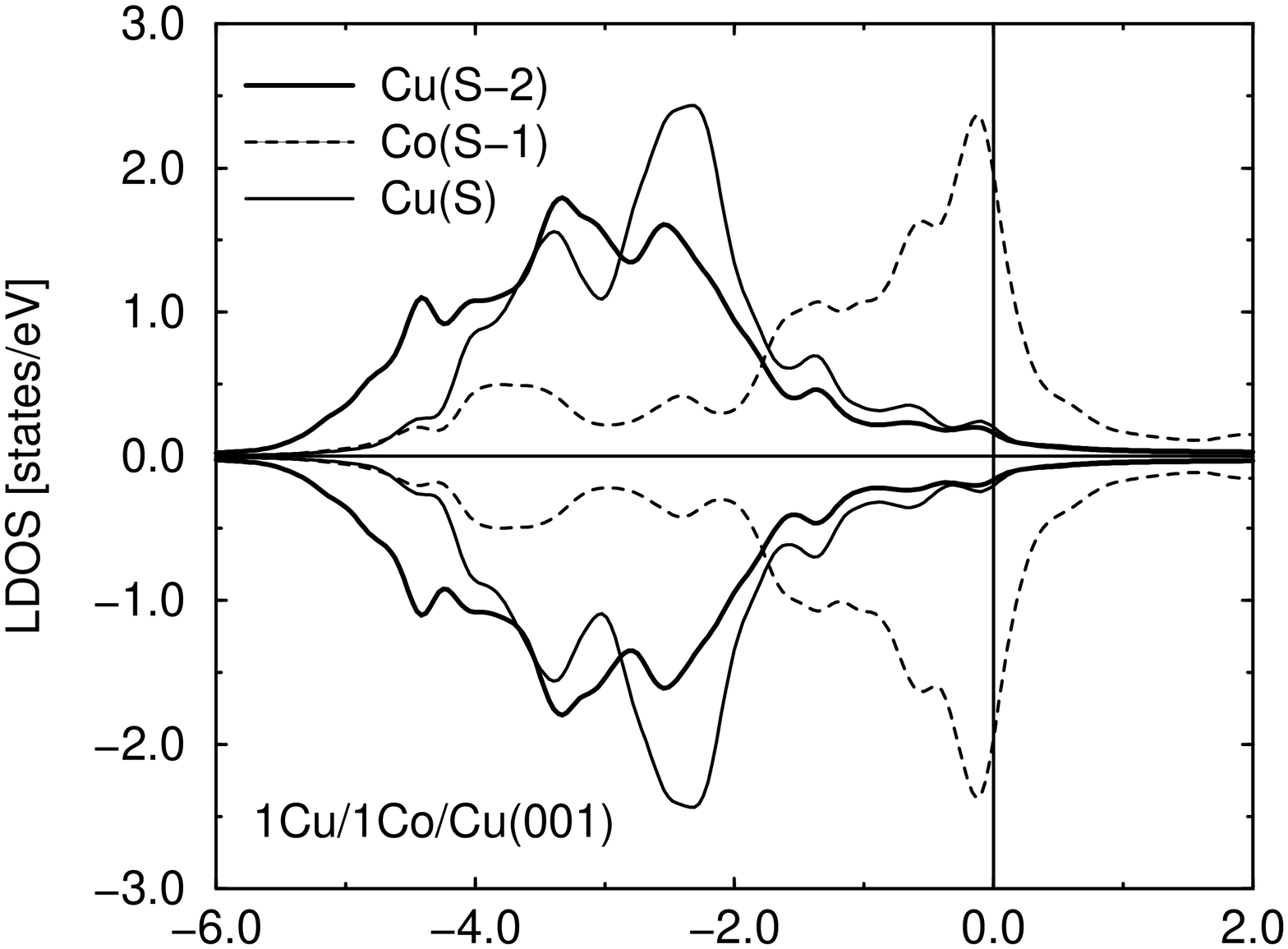,width=6.5cm,angle=270}}
      \vspace*{-0.8cm}
\centerline{\psfig{figure=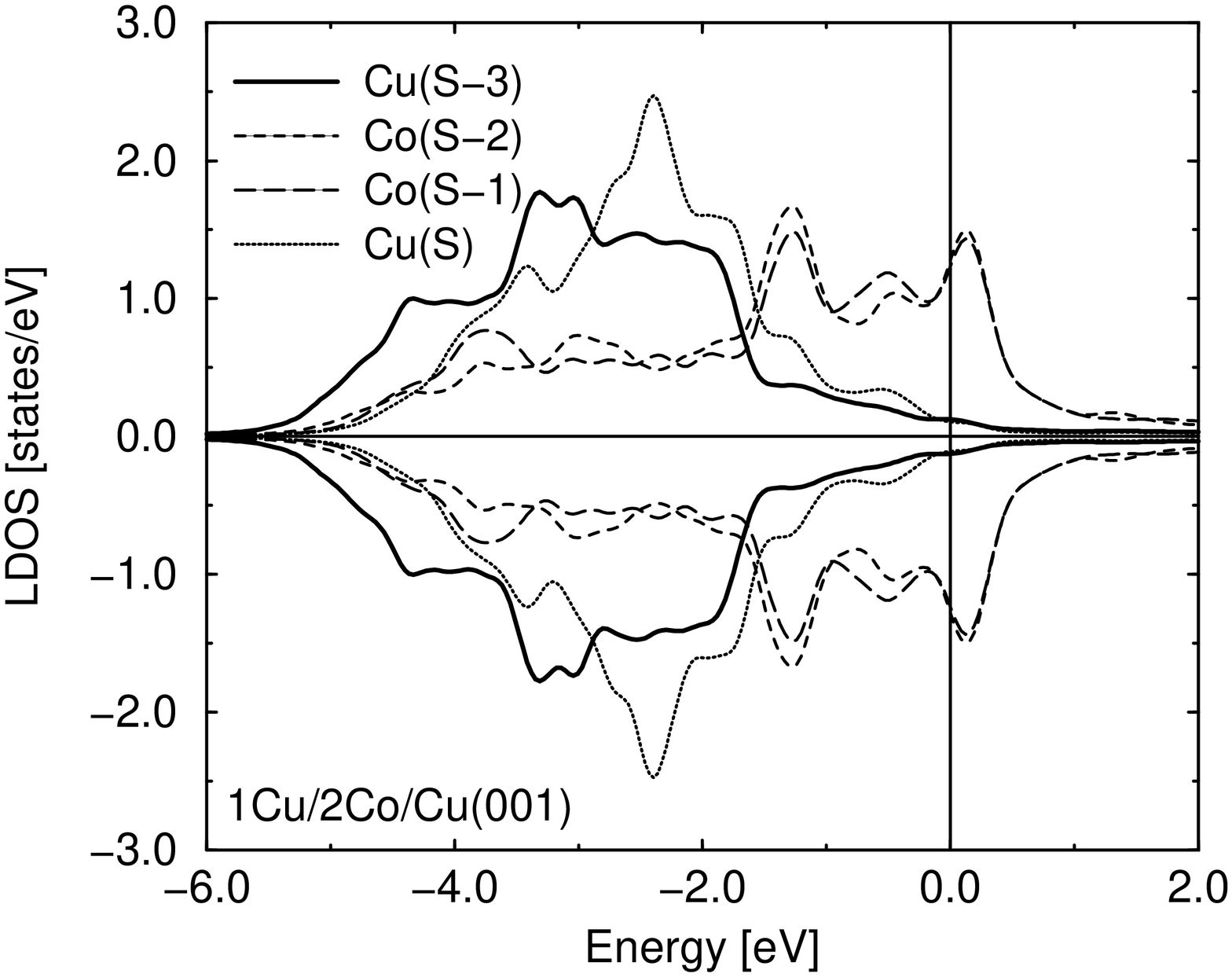,width=6.5cm,angle=270}}
      \vspace*{0.3cm}
   \caption{\label{ldosd_nm} Local density of states of the $d$-bands of the
     different atomic layers in the nonmagnetic
     systems. S, S-1, S-2, S-3 denote the surface and the subsequent
     subsurface layers respectively. We display the contribution from inside
     the muffin-tin spheres. The Co bands are represented with dashed and
     long-dashed lines, the Cu bands with a solid line. The calculated LDOS was
     broadened by a Gaussian with a width of $2\sigma=0.2$~eV.}
\end{figure}

The calculated electronic properties are consistent with the 
above discussed structural and energetic trends. Figure~\ref{ldosd_nm} 
shows the local density of states (LDOS) of
the $d$-bands of the adsorbate and substrate layers obtained from a
nonmagnetic calculation. For  1Co/Cu(001) the Co $d$-band 
is rather narrow, the LDOS at the Fermi level is very high
and the overlap with the copper $d$-band is small, which reflects
that the interaction between  Co and Cu is not very strong. 
For the 2Co/Cu(001) system the $d$-states of the surface and subsurface Co layers
overlap and their $d$-bands receive a substantial broadening.
At the same time the LDOS at the Fermi level is lowered. The broadening of the 
cobalt $d$-bands in 2Co/Cu(001) is an indication for the strong interaction
between the two cobalt layers. The same effect of broadening of the $d$-band of
Co is observed for 1Cu/2Co/Cu(001) compared to the Co $d$-band in 1Cu/1Co/Cu(001). 

\begin{figure}
\centering
\centerline{\psfig{figure=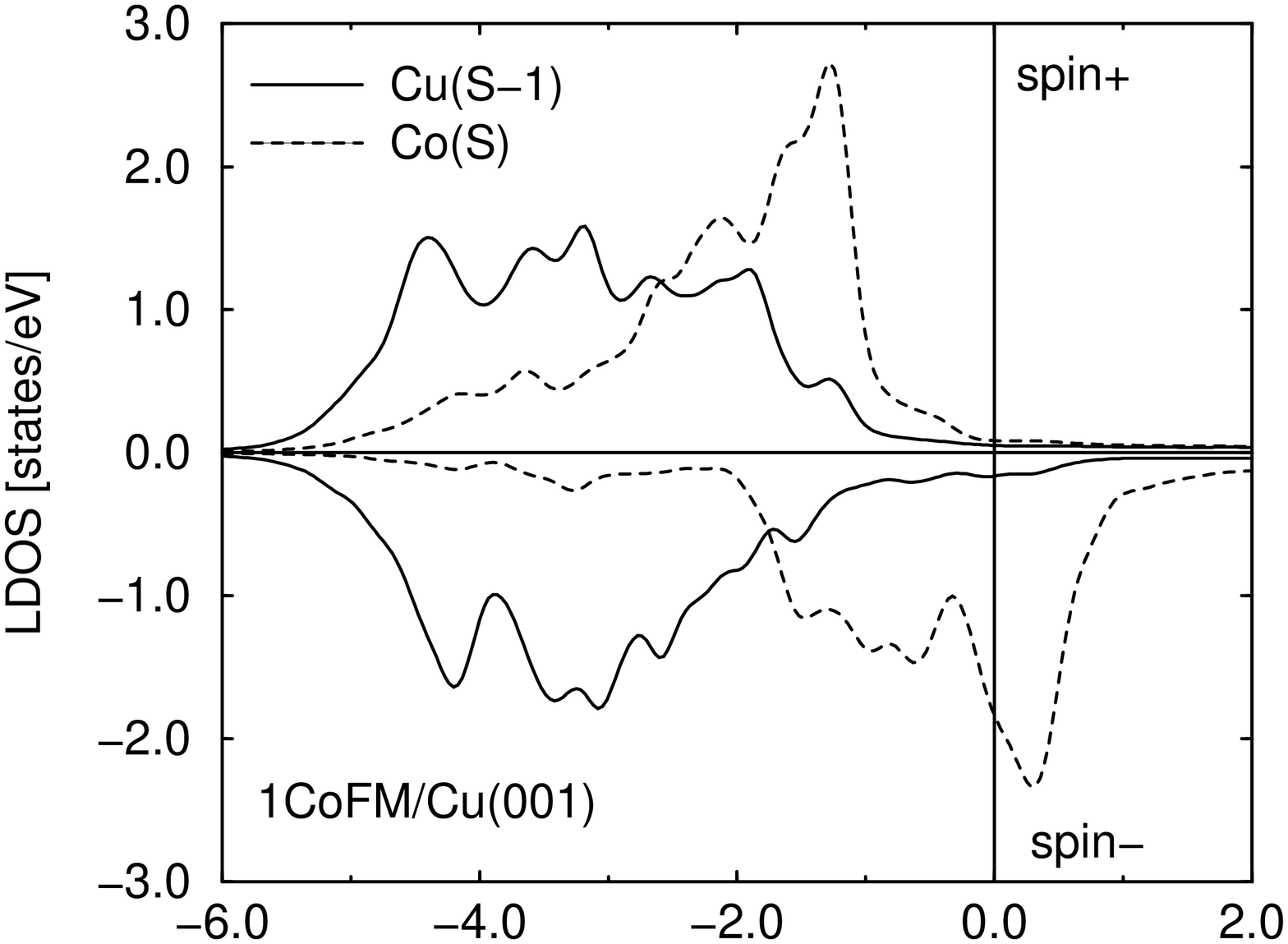,width=6.5cm,angle=270}}
      \vspace*{-0.8cm}
\centerline{\psfig{figure=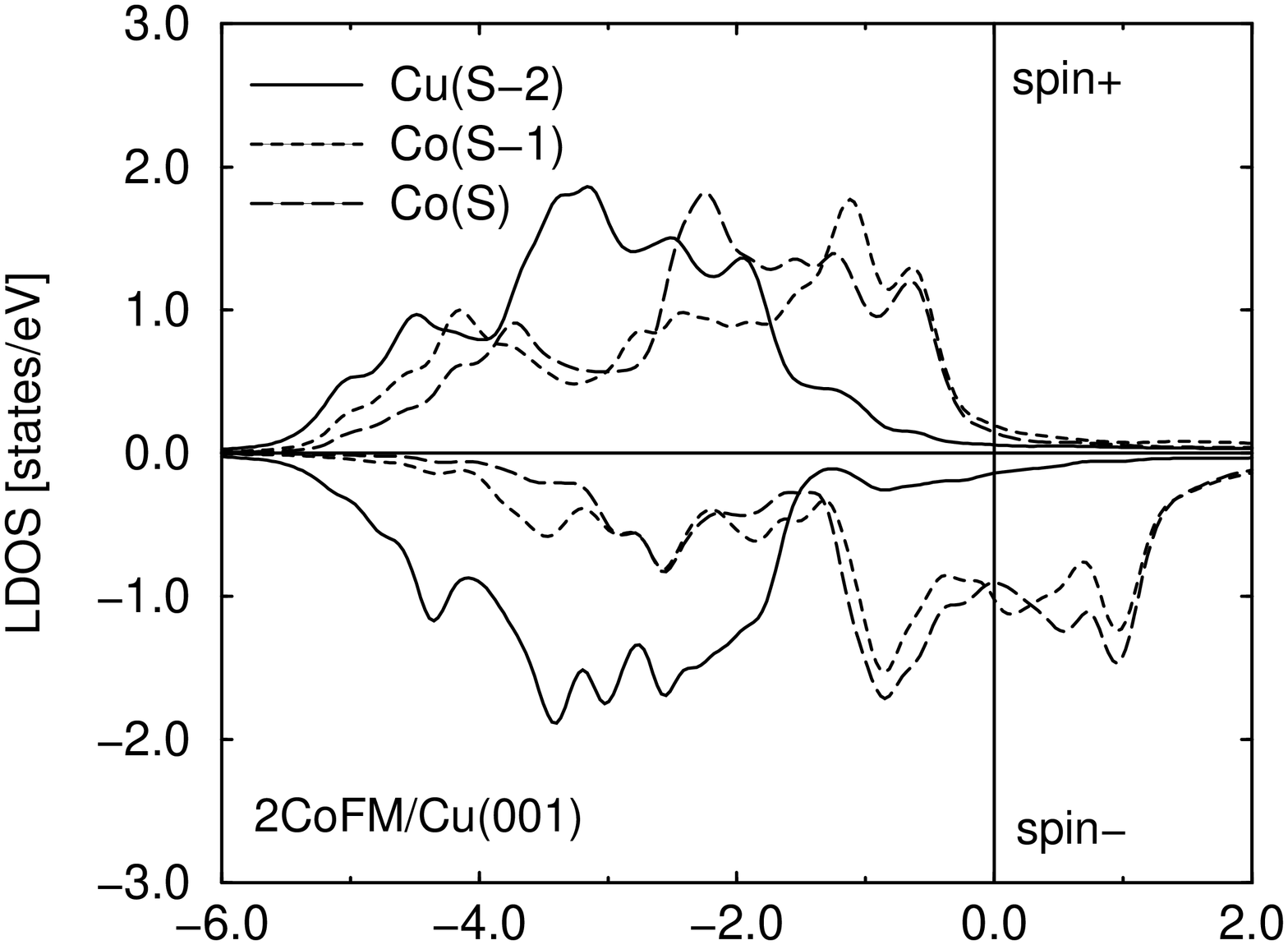,width=6.5cm,angle=270}}
      \vspace*{-0.8cm}
\centerline{\psfig{figure=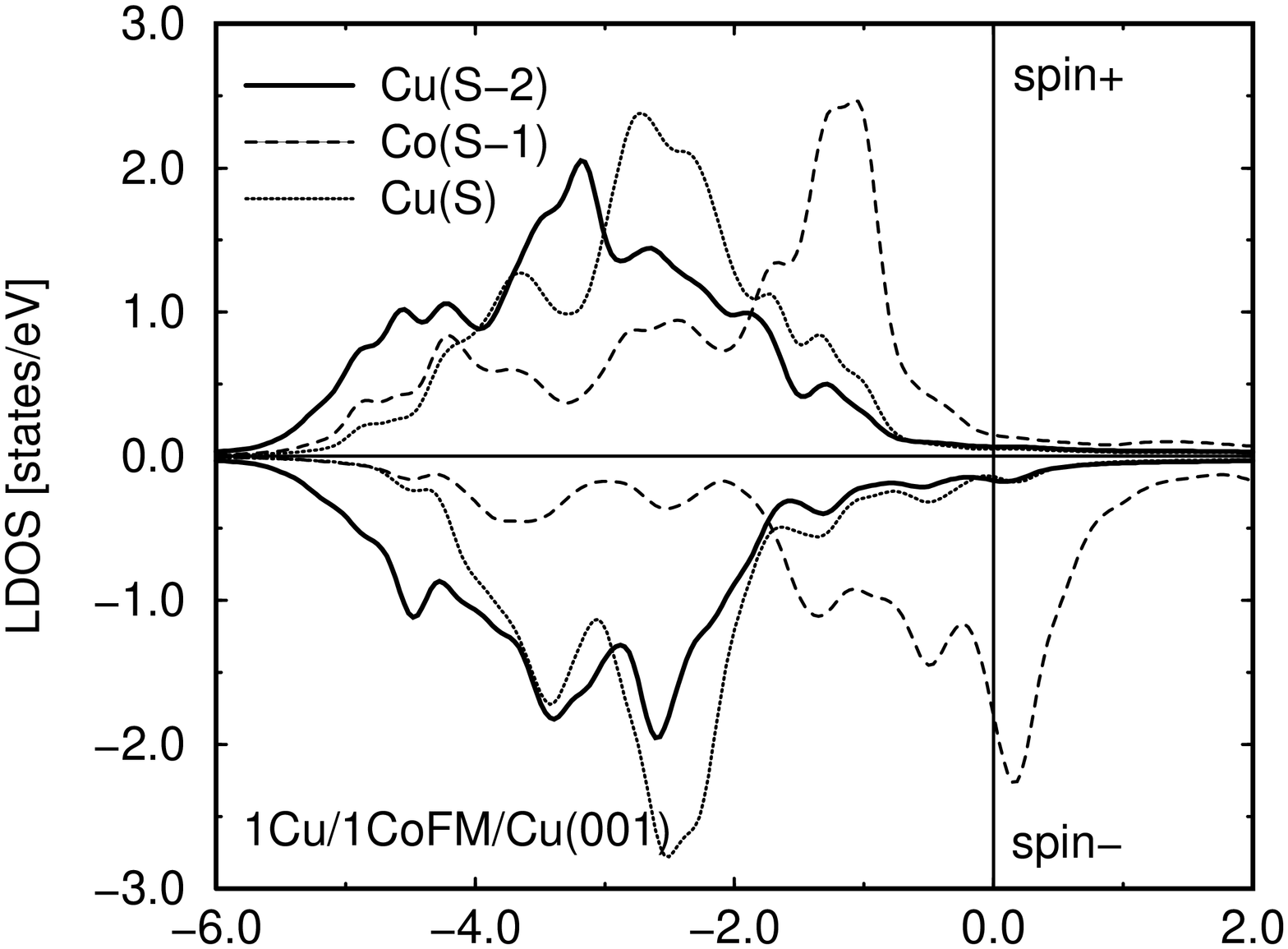,width=6.5cm,angle=270}} 
      \vspace*{-0.8cm}
\centerline{\psfig{figure=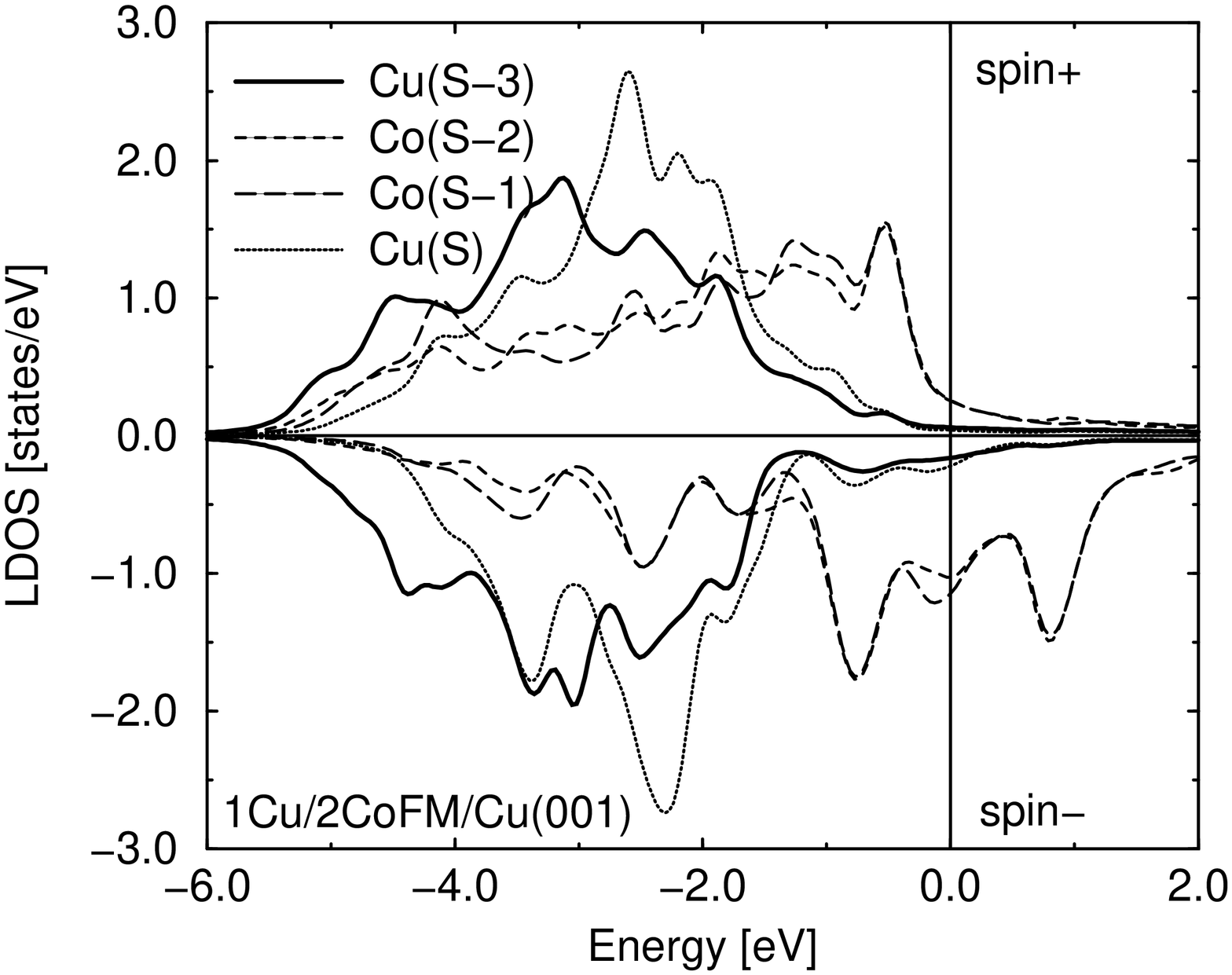,width=6.5cm,angle=270}}
      \vspace*{0.3cm}
   \caption{\label{ldosd_fm} LDOS of the $d$-bands of the
     different atomic layers in the ferromagnetic
     systems.S, S-1, S-2, S-3 denote the surface and the subsequent
     subsurface layers respectively. The contribution from inside
     the muffin-tin spheres is displayed. The Co bands are marked with dashed
     and long-dashed lines, the Cu bands with a solid line. The calculated
     LDOS was broadened by a Gaussian with a width of $2\sigma=0.2$~eV.}
\end{figure}

The layer-resolved LDOS of the $d$-bands for the ferromagnetic systems is
given in Fig.~\ref{ldosd_fm}. The majority band of Co is completely filled and
the minority band is only partly filled, reflecting the fact that Co is a strong
ferromagnet. For 1CoFM/Cu(001) and 1Cu/1CoFM/Cu(001) the Fermi level crosses
the minority $d$-band of cobalt almost at its maximum while for 2CoFM/Cu(001)
and 1Cu/2CoFM/Cu(001) the Fermi level lies in a dip of the Co $d$-bands. 
A ``harder'' electronic structure, i.e. lower density at the 
Fermi level is typically considered an indication for a more stable system.

Both majority and minority $d$-bands of copper are occupied and lie ca. 2~eV
below the Fermi level. Still the minority and majority $d$-band have a very
different structure with the majority band being broader in general. Actually
the band width correlates with the strength of interaction with the cobalt film:
While there is a substantial overlap between the majority $d$-bands of cobalt
and the substrate layer beneath or the capping layer above, the corresponding
minority bands have a very small overlap.   
\section{Comparison of Co/Cu(001) and Co/Cu(111)}\label{co111_co001}
Prior to our work Pedersen {\em et al.}~\cite{norskov} studied the growth of
Co on the (111) surface of Cu with STM and LMTO-calculations. The STM
measurements showed that the islands consist of several cobalt layers with the
lowest layer possibly growing subsurface. At elevated temperatures vacancy islands
formed in the terraces close to steps and the substrate
material etched from these holes covered the cobalt islands.

Comparing the LMTO calculations for nonmagnetic systems with
$[111]$-orientation\cite{norskov} with our FP-LAPW
results for the $[001]$-orientation we note 
that the general behavior is similar:
The systems are unstable against phase separation and
clustering. The energy gain from the separation of a monolayer film in a
bilayer island and a clean Cu surface is $\Delta E_{(001)} =
0.59$~eV/$(1\times 1)$-cell and $\Delta E_{(111)} = 0.39$~eV/$(1\times
1)$-cell respectively. The corresponding energy gain for the capped 
systems is
$\Delta E_{(100)} = 0.31$~eV/$(1\times 1)$-cell and $\Delta E_{(111)} =
0.18$~eV/$(1\times 1)$-cell.

The segregation of Cu onto the surface lowers the energy of the system for both
orientations: for monolayer coverages $\Delta E_{\rm (100)} =
0.49$~eV/$(1\times 1)$-cell and $\Delta E_{(111)} = 0.30$~eV/$(1\times
1)$-cell; for 2ML of Co $\Delta E_{(100)} = 0.42$~eV/$(1\times 1)$-cell and
$\Delta E_{(111)} = 0.20$~eV/$(1\times 1)$-cell.

Still in all cases the energy gain is lower for the (111) surface
than the (100) surface. This trend reflects the difference in 
coordination numbers: In a bond cutting model of metallic bonding 
the energy of an atom roughly scales as the square root of the local
coordination~\cite{Spanjaard90,methf2}. 
Adsorption of a Co layer or of a Cu capping layer implies
a change of coordination of the atoms in the added layer
from 4 to 8 for the $[100]$- and 
from 6 to 9 for the  $[111]$-orientation. And for the atoms
in the layer, which after adsorption becomes the second layer,
the coordination changes from 8 to 12 for the $[100]$- and 
from 9 to 12 for the  $[111]$-orientation.
Thus, the energy gain is smaller for the $[111]$-orientation
than for the $[100]$-orientation.

\section{Summary}\label{Summary}
In summary, we identify a bilayer cobalt island covered by copper as the
lowest energy configuration. However, growth is ruled by kinetics. 
Therefore, it is to be expected that under realistic conditions metastable
structures may exist at surfaces, and some examples were 
identified in this paper. Total-energy considerations show that
the $(1\times 1)$-film tends to separate into areas of bilayer cobalt 
islands and clean copper surface, and this is indeed in line with 
experimental observations of bilayer growth~\cite{lit,kir1,kir5}. Our 
total energy and electron density of states results show that the 
stability of the bilayer film is due to the fact that the Co atoms prefer to
attain a high coordination of alike atoms. A consequence of this is the very strong 
contraction of the interlayer distance between the two cobalt layers and
the substantial broadening of the cobalt $d$-bands in the adsorbed Co-bilayer
as compared to the $d$-band of a single Co adlayer.

The segregation of substrate material onto the Co adlayer
results in a substantial energy gain $0.5$~eV/$(1\times 1)$-cell).
We also studied a two-layer
surface alloy of Co and Cu with a  $c(2\times 2)$ periodicity.
This is found to be energetically less favorable than a separation into
Cu-capped Co bilayer-adsorbates, but at strained regions of
the surface this surface alloy may be stabilized.
Indeed, a $c(2\times 2)$ surface structure was observed in recent 
combined STM and RHEED experiments~\cite{nouv}. 

Generally, the ferromagnetically ordered systems are lower in energy than
the nonmagnetic, but the relative stability of different configurations remains
qualitatively unchanged by magnetism and the structural trends are well
described by the nonmagnetic systems. 

For low coverages ($\Theta < 0.25$~ML) we also find that cobalt may 
adsorb substitutionally~\cite{prb1,prb2}. However, with increasing coverage the
substitutional adsorption becomes energetically unfavorable compared to the
formation of compact islands.
\acknowledgements
We gratefully acknowledge discussions with P. Kratzer,
P. Ruggerone, and P. M. Marcus.
We also thank M. Farle and K. Baberschke for
stimulating discussions. The work was supported by the DFG through SFB290.
We thank A. Chaka for a careful reading of the manuscript.
\section{Appendix}
For several systems we performed calculations with the LDA~\cite{pw} 
and with the  GGA~\cite{pbe96}. The formation energies for 1Co/Cu(001) (NM and FM),
2Co/Cu(001) (NM and FM), 1Cu/1Co/Cu(001), and 1Cu/2Co/Cu(001) are given in
Table~\ref{ggalda_tab}. The lateral parameter was set to the lattice constants
of copper obtained within the LDA and GGA approach, respectively. The LDA
value $3.55$~{\AA} is $1.7\%$ smaller than the measured one, $3.61$~{\AA},
while with the GGA the lattice parameter ($3.65$~{\AA})
is  $1.1\%$ bigger than the experimental value (zero-point vibrations 
are neglected in the theory).
\begin{table}
\begin{tabular}[h]{ c c c }
System   & $E^{\rm f}_{\rm LDA}$ & $E^{\rm f}_{\rm GGA}$      \\
\hline
Cu(001)         & 0.78     & 0.61    \\
\hline
1Co/Cu(001) NM  & 1.75     & 1.47    \\
1Co/Cu(001) FM  & 1.54     & 1.22    \\
\hline
2Co/Cu(001 NM  &   1.55    & 1.27      \\
2Co/Cu(001 FM  &  1.48     & 1.11  \\
\hline
1Cu/1Co/Cu(001) NM &   1.26 & 0.99      \\
\hline
1Cu/2Co/Cu(001) NM &  1.13  & 0.85      \\
\end{tabular}
\centerline{
      \parbox{ 13.0cm}{
        \caption 
{\small{The formation energies $E^{\rm f}_{\rm LDA}$ and $E^{\rm f}_{\rm GGA}$ 
    of the different configurations calculated within LDA and GGA,
    respectively, given in [eV/$(1\times 1)$-cell]. The lateral parameter is
    set to the corresponding (LDA or GGA) equilibrium lattice constant of
    copper.}\label{ggalda_tab}}}}  
\end{table}

The formation energies obtained with the GGA are generally lower than the LDA
results and the differences are between $0.2$ and $0.3$~eV/$(1\times
)$-cell. This effect was also  observed previously for the clean copper
surface~\cite{boisvert97}. Yet the trends between the different configurations
remain unchanged. For example the energy gain from the separation of a
monolayer cobalt film in a bilayer cobalt island and clean Cu(001)-surface is
$0.60$~eV/$(1\times 1)$-cell (LDA) and $0.53$~eV/$(1\times 1)$-cell (GGA) for
the nonmagnetic systems and $0.41$~eV/$(1\times 1)$-cell (LDA) and
$0.36$~eV/$(1\times 1)$-cell (GGA) for the ferromagnetically ordered
systems. The equilibrium configuration of clean copper surface with bilayer
cobalt islands, covered by copper (see Fig.~\ref{fig:conf1.0}(f)), is by
$0.79$~eV/$(1\times 1)$-cell (LDA) and by $0.74$~eV/$(1\times 1)$-cell (GGA)
more favorable than 1Co/Cu(001) in Fig.~\ref{fig:conf1.0}(a).

A structural optimization was performed for all systems listed in
Table~\ref{ggalda_tab} using both approaches, LDA and GGA. No
noticeable differences were obtained except for the systems, containing a
bilayer cobalt film, where the contraction of the distance between the two
cobalt layers was slightly stronger with the GGA, e.g. for 2Co/Cu(001) $\Delta
d_{\rm Co-Co}^{\rm LDA}/d_0^{\rm LDA}=17\,\%$ and $\Delta d_{\rm Co-Co}^{\rm
  GGA}/d_0^{\rm GGA}=18.6\,\%$. However, these minor differences do not alter
the discussion in Section~\ref{struct}. 

The larger lateral parameter in GGA produces a substantial
enhancement of the magnetic moments, e.g. the surface magnetic moment of
cobalt in  1Co/Cu(001) changes from $1.71\,\mu_B$ (LSDA) to $1.86,\mu_B$ (GGA)
and in 2Co/Cu(001)  from $M_{\rm Co(S)}^{\rm LSDA} = 1.71\,\mu_B$ and $M_{\rm
  Co(S-1)}^{\rm LSDA} = 1.47\,\mu_B$ to $M_{\rm Co(S)}^{\rm GGA} =
1.81\,\mu_B$ and $M_{\rm Co(S-1)}^{\rm GGA} = 1.64\,\mu_B$. This result is not
surprising and is in line with the changes of the magnetic moment for fcc
cobalt bulk from $1.52\,\mu_B$ (LSDA) to $1.69\,\mu_B$ (GGA). 

In conclusion, both approximations of the exchange-correlation 
potential, LDA and GGA, lead to the same results for the structural,
energetic, and magnetic properties of the configurations studied in this work.


\newpage
%

%


\begin{references}
\bibitem{bil-c2x2}     The calculations for bilayer $c(2 \times 2)$-alloy were
                       performed with the same cutoff parameters as for the
                       $(1\times 1)$-structures (see Section~\ref{theory} and
                       15 $\mathbf{k}_{\parallel}$-points in the irreducible
                       wedge of the Brillouin-zone.
\bibitem{co_fct}       For fct cobalt with $a_{\parallel}=a_{\rm Cu}$ we
                       obtained the following distances in $z$-direction:
                       $a_z^{\rm NM}=3.23$~{\AA} and $a_z^{\rm
                           FM}=3.31$~{\AA}.
\bibitem{e_co_fct}     We note here that the formation energy of
                       Co(001) was also calculated with respect to fct cobalt
                       bulk in order to separate the elastic from the electronic
                       effects.
\bibitem{umls}         The fact that the equilibrium lattice constant of a free-
                       standing Co monolayer is noticeably smaller than the 
                       lattice constant of Co (or Cu) bulk reflects that the low 
                       coordination of atoms in the free-standing monolayer
                       leads to a stronger bonding between the cobalt atoms and 
                       thus shorter bond lengths. This result follows Pauling's 
                       description of the relation between coordination, bond 
                       strength, and bond length.
\end{references}
\end{document}